\magnification=\magstep1
\overfullrule=0pt
\parskip=6pt
\baselineskip=15pt
\headline={\ifnum\pageno>1 \hss \number\pageno\ \hss \else\hfill \fi}
\pageno=1
\nopagenumbers
\hbadness=1000000
\vbadness=1000000

\input epsf


\vskip 25mm
\vskip 25mm
\vskip 25mm
\vskip 25mm
\vskip 25mm

\centerline{\bf  POINCARE POLINOMIALS OF HYPERBOLIC LIE ALGEBRAS OF RANK THREE }
\vskip 15mm

\centerline{\bf M. Gungormez}
\centerline{Dept. Physics, Fac. Science, Istanbul Tech. Univ.}
\centerline{80626, Maslak, Istanbul, Turkey }
\centerline{e-mail: gungorm@itu.edu.tr}

\medskip

\centerline{\bf{Abstract}}

In view of a previous work, we explicitly give the Poincare polinomials
of 19 Hyperbolic Lie algebras of rank 3. It is seen that every one of these 
polinomials is expressed as the ratio of Poincare polinomial of $B_3$ Lie
algebra and a polinomial of finite degree.

\vskip 15mm
\vskip 15mm
\medskip

\hfill\eject

\vskip 3mm
\noindent {\bf{I.\ INTRODUCTION }}
\vskip 3mm

In a previous work {\bf[1]}, we have shown that Poincare polinomials of Hyperbolic 
Lie algebras are expressed as the ratio of Poincare polinomial of a properly chosen
finite Lie algebra and a polinomial of some finite degree. In view of the fact 
that there is only a finite number of this type of Lie algebras {\bf[2]}, we now 
present the case for rank 3 for which we have 19 different ones. It is interesting 
to note that $B_3$ Lie algebra plays a central role here. 

For a brief summary of our method which is presented in {\bf[1]} in detail, we note that 
Poincare polinomials{\bf[3]} could be defined as being in relation with Weyl groups. 
Let us consider 
$$ P(t) \equiv \sum_{n=0}^\infty  q(n) \ t^n $$
as Poincare polinomial of a Kac-Moody Lie algebra {\bf[4]} $G_r$ of rank  r where t is
some indeterminate. q(n) here is the number of Weyl group elements which are defined
as products of n simple Weyl reflections. Let $ {\cal W}(G_r)$ be the Weyl group, $\rho$ 
Weyl vector and $\alpha_i$'s be the simple rots of $G_r$ where $i=1,2 \dots r.$ As is given 
in {\bf[1]}, our main objective to obtain q(n) 
is then simply the statement that
$$ \gamma \equiv \rho-\Sigma(\rho) \eqno(I.1)$$
is unique for each and every element $\Sigma \in {\cal W}(G_r)$. Note that $\gamma$ is 
by definition an element of positive root lattice. Beside the set $ \Gamma^0 $ which
consists of zero root solely, the set
$$\Gamma^1 \equiv \{ \alpha_1, \alpha_2, \dots , \alpha_r \} $$
contains the first r+1 solutions of (I.1) and $ \Gamma^1$ corresponds by definition to
simple Weyl reflections. All these are related with the fact that $ q(0) = 1 $ and also 
$ q(1) = r$ as a result of
$$ q(0) = dim \Gamma^0 $$
\noindent and
$$ q(1) = dim \Gamma^1 $$
where $ dim \Gamma^n$ means the number of elements for any $\Gamma^n$ which corresponds 
to Weyl group elements consisting of products of n simple Weyl reflections. One can readly 
see that all the sets $\Gamma^n $ of solutions of (I.1) are to be constructed recursively 
from $\Gamma^1$. Our results will be presented for 19 Hyperbolic Lie algebras of rank 3 in 
the next section.

\vskip 3mm
\noindent {\bf{II.\ RANK THREE CLASSIFICATION }}
\vskip 3mm

Let $ {\cal H}^s $ be one of the 19 Hyperbolic Lie algebra of rank 3 as given in {\bf[2]}
\break where $s=1,2, \dots 19.$ Our main observation here is
$$ P^s(t) \equiv { P_{B_3}(t) \over Q_s(t) } \eqno(II.1) $$
for Poincare polinomial $P^s(t)$ of $ {\cal H}^s $ where $ P_{B_3}(t) $ is the Poincare polinomial of $B_3$ Lie algebra and $Q_s(t)$ is a polinomial of some finite degree of 
indeterminate t. $B_3$ here is defined by the diagram

\epsfxsize=4cm
\centerline{\epsfbox{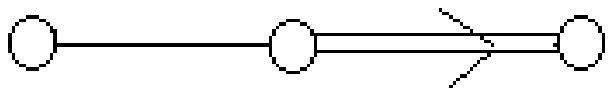}}

\noindent and the polinomials $Q_s(t)$ will be given as in the following:
 
\epsfxsize=13cm
\leftline{\epsfbox{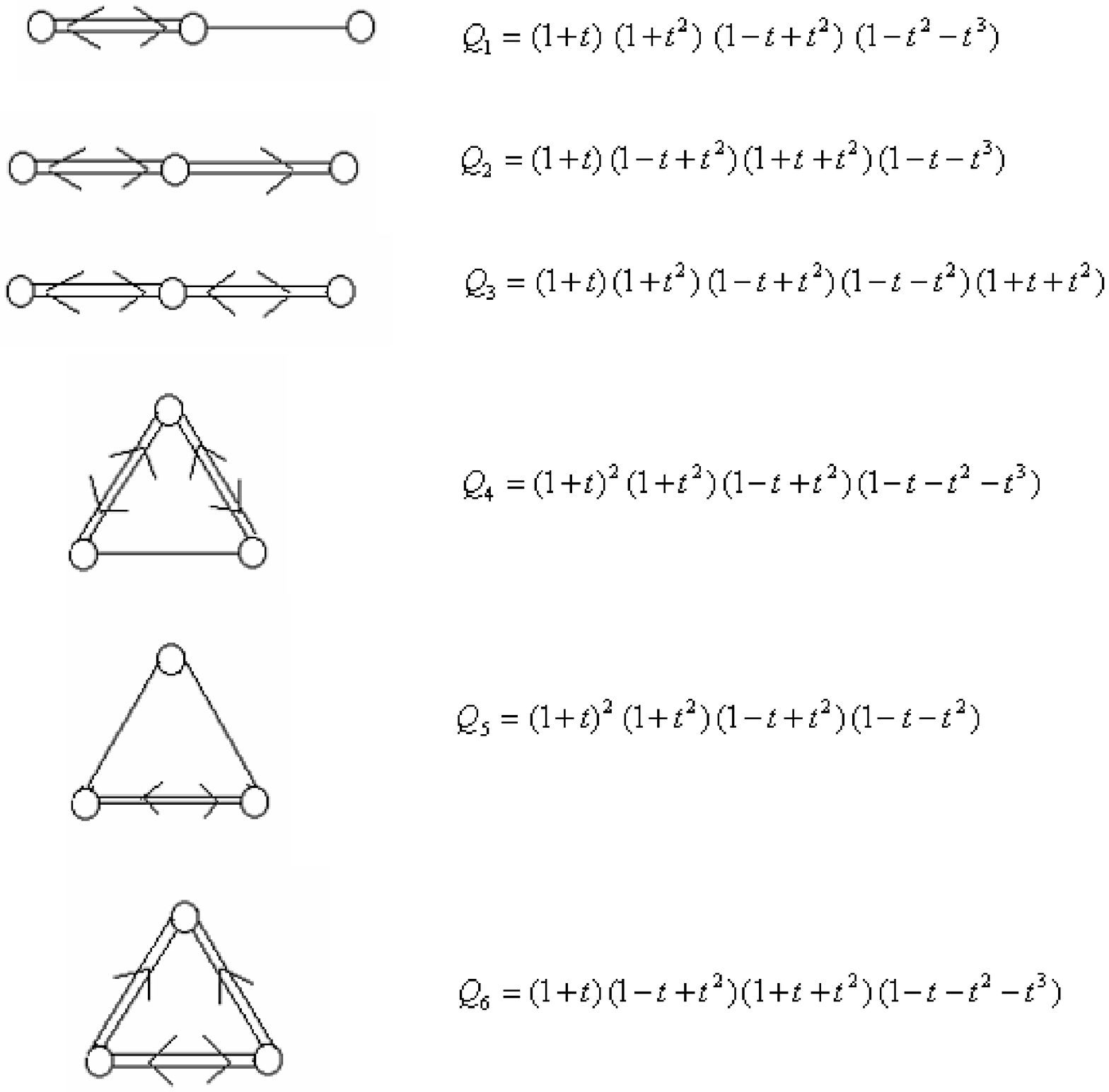}}

\epsfxsize=13cm
\leftline{\epsfbox{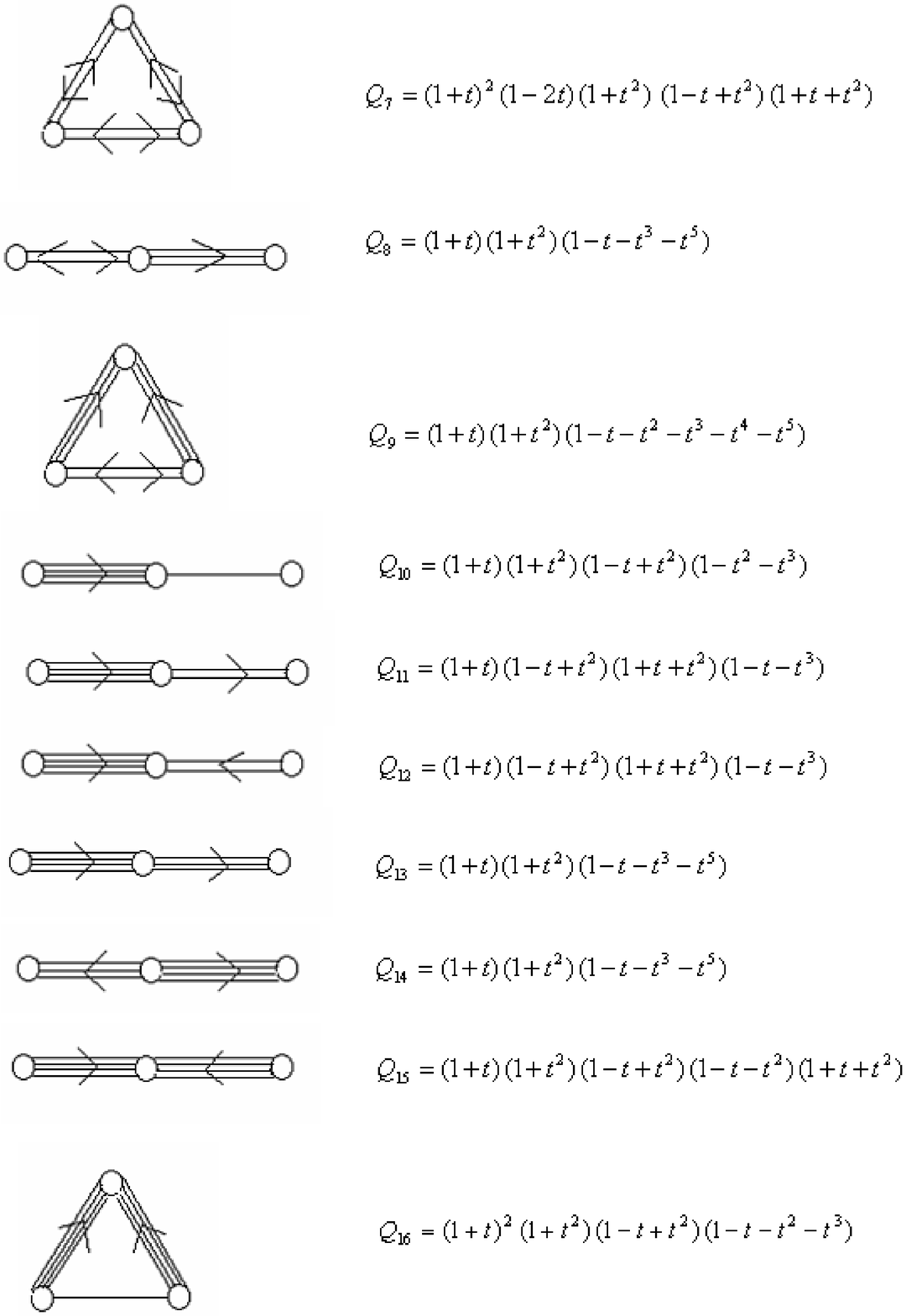}}

\epsfxsize=13cm
\leftline{\epsfbox{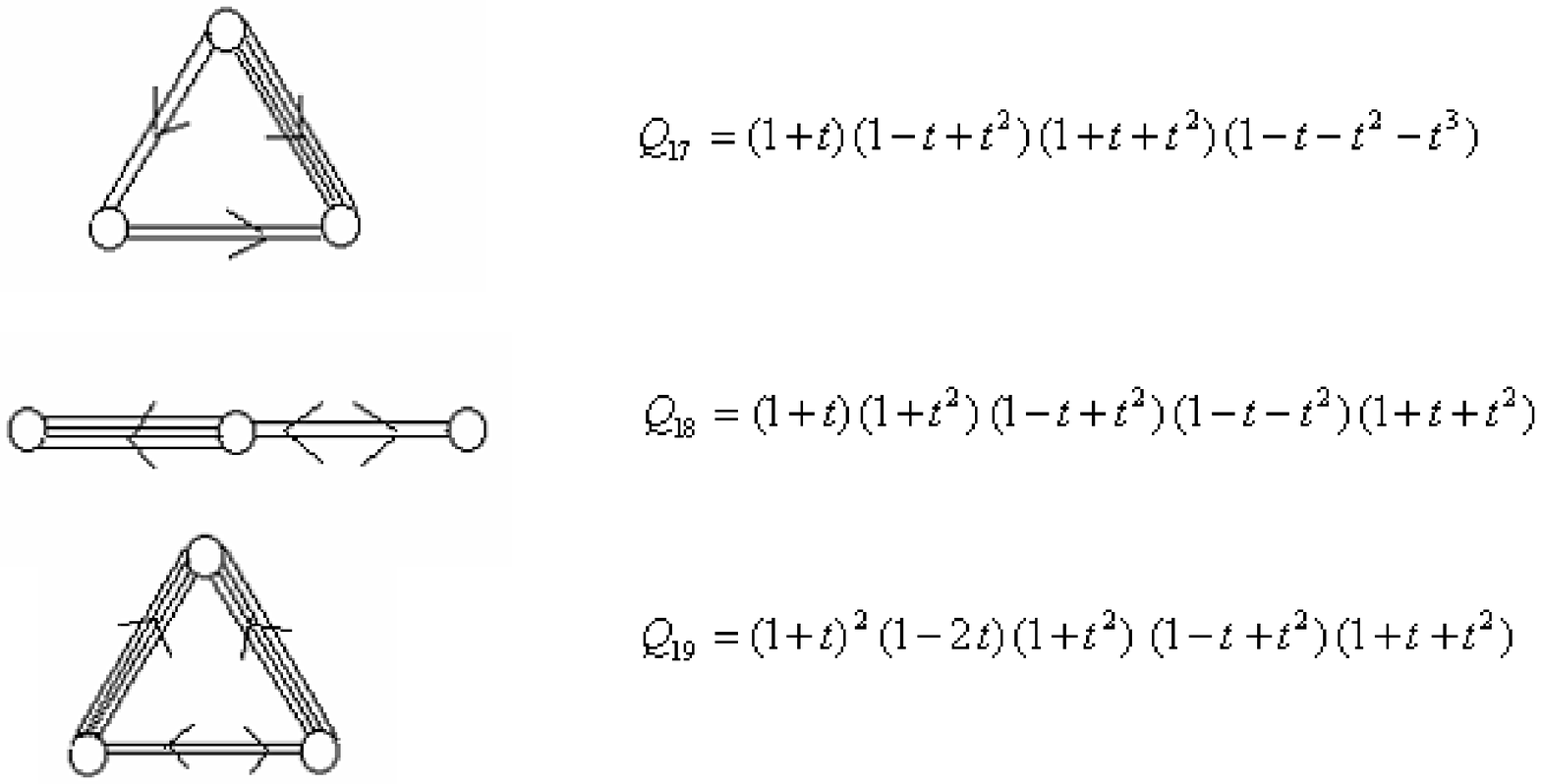}}

Let us finally note that one could ask, beside (I.1), if there is an alternative form,
for instance
$$ P^s(t) \equiv { P_{A_3}(t) \over R_s(t) } \eqno(II.2) $$

\noindent where $P(A_3)$ is Poincare polinomial of $A_3$ Lie algebra and $R_s(t)$'s are 
some other set of polinomials of some finite order. We also studied this problem and find 
that one can find solutions except s=8,9,13,14 hence $B_3$ is unique for this instance.

\vskip 3mm

\noindent {\bf{ACKNOWLEDGEMENT}}

The author would like thank H.R.Karadayi for his continous support during this work.

\vskip3mm
\noindent{\bf{REFERENCES}}
\vskip3mm

\item [1] H.R.Karadayi and M.Gungormez, On Poincare Polinomials of Hyperbolic Lie 

\item \ \ \ \ \ Algebras(submitted for publication)

\item [2] C. Saclioglu, J.Phys. A:Math.Gen(1989) 3753-3769

\item [3] J.E.Humphreys, Reflection Groups and Coxeter Groups, 

\item \ \ \ \ \ Cambridge University Press, 1990

\item [4] V.Kac, Infinite Dimensional Lie Algebras

\item \ \ \ \ Cambridge University Press, 1982

\end